\documentclass[12pt]{article}

\usepackage[russian]{babel}               
\begin{document}
\begin{center}
{\large\bf Vector Theory of Gravity} \vskip 0.3 true in
{\large V. N. Borodikhin} \vskip 0.3 true in {\it
Omsk State University, pr. Mira 55a, Omsk, Russia}
\end{center}
\begin{abstract}%
We proposed a gravitation theory based on an analogy with electrodynamics on the basis of
a vector field. For the first time, to calculate the basic gravitational effects in the framework of a vector
theory of gravity, we use a Lagrangian written with gravitational radiation neglected and generalized to the
case of ultra-relativistic speeds. This allows us to accurately calculate the values of all three major gravity
experiments: the values of the perihelion shift of Mercury, the light deflection angle in the gravity field of the
Sun and the value of radar echo delay. The calculated values coincide with the observed ones. It is shown
that, in this theory, there exists a model of an expanding Universe.
\end{abstract}

\vskip 0.2 true in e-mail: borodikhin@inbox.ru

\section{Introduction}

In this paper we make an attempt to describe
the gravitational phenomena using the vector field
approximation in Minkowski space. Some attempts
to describe the gravity using vector models were made
previously
 \cite{0,3,2}, but a number of difficulties arose in
this approach. The main problems are the absence of
light deflection in the gravitational field and an incorrect
value of the anomalous precession of Mercury's
perihelion \cite{3}.

It should be noted, however, that all the calculations
of these effects in the framework of a vector
theory of gravity were performed either ignoring corrections
related to the vector field or in the Newtonian
approximation.

We use the Lagrangian of a vector gravitational
field written neglecting gravitational radiation. Since
radiation is due to acceleration, the assumption of a
low speed should not be necessary. Thus it is possible
to generalize the Lagrangian without radiation for
the case of ultra-relativistic speeds, and it is done
here. (Such a generalization for electromagnetic field
theory has been described in \cite {5}).

This allows us
to accurately calculate the values of all three major
gravity experiments: the perihelion shift of Mercury,
light deflection by the gravity field of the Sun and
radar echo delay. The calculated values coincide with
the observed ones \cite{3,1}.
Such a Lagrangian written
generalized to large velocities is used in a vector theory
of gravity for the first time.
It is also shown that cosmologicalmodels of vector
gravity theory, neglecting the vector potential, are
equivalent to the standard flat, closed and hyperbolic
Universe models. These cases correspond to the
simple Newton-type picture with elliptic, parabolic
and hyperbolic matter motion depending on the initial
velocity. The cause of the initial velocity distribution
is unknown.

\section{The general model}

We will connect the gravitational field with the 4-potential $A^i=(\varphi, c\vec A)$, where
$\varphi$- is the usual scalar potential and $\vec A$ is a vector potential, and $c$ is the speed of light.
The Lagrangian of the gravitational field with
account for matter has the form

\begin{equation}\label{4}
\L=-A_ij^i+\frac{1}{16\pi\gamma}G_{ik}G^{ik},
\end{equation}

where $\gamma$ is the gravitation constant,
$j^i=\mu\frac{1}{c}\frac{dx^i}{dt}$ is
the mass current density vector
, $\mu$- is the mass density
of bodies, and
$G_{ik}=\frac{\partial A_k}{\partial x^i}-
\frac{\partial A_i}{\partial x^k} $ is the antisymmetric tensor of the gravitation field.

The first term describes interaction of the field and
matter, the second one characterizes the field without
particles.
As a result, we get the gravitational field equations

\begin{equation}\label{5}
\frac{\partial G^{ik}}{\partial x^k}=4\pi\gamma j^i.
\end{equation}

In the stationary case, from (\ref{5}) we obtain an equation
for the scalar potential:

\begin{equation}\label{7}
\triangle\varphi=4\pi\gamma\mu.
\end{equation}

The solution of (\ref{7}) has the form

\begin{equation}\label{8}
\varphi=-\gamma\int\frac{\mu}{r}dV.
\end{equation}

The potential of a single particle of mass m $\varphi=-\frac{\gamma m}{r}$.
Consequently the force acting in this field on
another particle of mass $m^{\prime}$ is

\begin{equation}\label{9}
F=-\frac{\gamma m m^{\prime}}{r^2},
\end{equation}

(\ref{9}) which is the Newton law of gravity. The negative sign
in this expression is caused by the positive sign of the
second term in the Lagrangian (\ref{4}), contrary to the
electromagnetic field Lagrangian \cite{7}.

Let us consider the field of the vector potential created
by matter particles performing motion in a finite
region od space with finite momenta. The motion of
this kind can be considered to be stationary. Let us
write down an equation for the time-averaged vector
field, depending only on spatial variables.

From (\ref{5}) we obtain:

\begin{equation}\label{10}
\triangle\overline{\vec A}=4\pi\gamma\overline{\vec j} ,
\end{equation}

whence it follows

\begin{equation}\label{11}
\overline{\vec A}= -\frac{\gamma}{c^2}\int\frac{\overline{\vec j}}{r}dV .
\end{equation}

The overline denotes a time average. This field can be
called cyclic. The field induction is

\begin{equation}\label{12}
\overline{\vec C}=rot\overline{\vec A}= -\gamma\int\frac{[\overline{\vec j}\vec r]}{r^3}dV=-\gamma,
\frac{[\overline{\vec p}\vec r]}{r^3},
\end{equation}

where $\vec p$ is the particle momentum and the square
brackets denote a vector product.

Thus two moving particles experience (in addition
to the mutual gravitational attraction) a cyclic force.
The latter can be attractive or repulsive, depending on
the relative direction of the particle velocities.

\section{Basic gravitational experiments}

Consider in this approach the calculation of the
main gravitational experiments: Mercury's perihelion
shift, light deflection in the gravitational field and the
radio signal delay.
From (\ref{4}) the Lagrangian of a body
of mass $m$ moving with the velocity $v$ is

\begin{equation}\label{ex1}
L=-mc^2\sqrt{(1-\frac{v^2}{c^2})}-m\varphi+m\vec v \vec A,
\end{equation}

where $\varphi$ is the scalar potential of the gravitational field, $\vec A$ is the vector
 potential of the cyclic field.
The solution of Eqs. (\ref{5}) in the general case is expressed in
terms of retarded potentials:

\begin{equation}\label{14-1}
\varphi=-\gamma\int\frac{\mu_{t-R/c}}{R}dV
\end{equation}

\begin{equation}\label{14-2}
\vec A= -\frac{\gamma}{c^2}\int\frac{\vec j_{t-R/c}}{R}dV ,
\end{equation}

where R is the distance from the volume element dV
to the point where the potential is sought for.
Expanding the scalar retarded potential up to the
second order with respect to the small parameter
$R/c$ and restricting ourselves to the first order for the
vector potential, let us insert the calculated potentials
into (\ref{ex1}).
Excluding the motion of the system as a
whole, we find the final result as the second-order
Lagrange function:

\begin{equation} \label{ex2}
L=\frac{m_1 v^2}{2}+\frac{m_2 v^2}{2}+\frac{\gamma m_1 m_2}{r}+\frac{\gamma m_1 m_2 v^2}{c^2 r}.
\end{equation}

For the system energy we can write:

\begin{equation} \label{ex3}
E=E_0-\frac{\gamma M m}{r}-\frac{\gamma MJ^2}{mc^2r^3}=E_0-V,
\end{equation}

where the velocity $\vec v=r\frac{d\psi}{dt}$ is expressed in terms
of the angular moment
 $J=mr^2\frac{d\psi}{dt}$, and $\psi$ is the angle; $M=m_1$, $m=m_2$.
It is convenient to calculate the perihelion shift
and the light deflection in a gravitation field using the
Runge–Lenz vector. For the first time this vector was
used for calculating the general-relativity corrections
in \cite{4}.

\begin{equation} \label{ex4}
\vec X=\vec v\times\vec J-\gamma Mm\vec e_r,
\end{equation}

where $\vec e_r$ -is a unit vector in the $r$ - direction.
The time derivative of the Runge-Lenz vector is

\begin{equation} \label{ex5}
\frac{d\vec X}{dt}=(r^2\frac{\partial V}{\partial r}-\gamma Mm)\frac{d\vec e_r}{dt}=
\Bigl(\frac{3\gamma MJ^2}{mr^2c^2}\Bigr)\frac{d\psi}{dt}\vec e_{\psi}.
\end{equation}

The direction of $\vec X$ changes with the angular velocity:

\begin{equation} \label{ex6}
\vec\omega=\frac{\vec X\times\vec{\dot X}}{\vec X^2}=
\Bigl(\frac{3\gamma MJ^2}{mr^2c^2X^2}\Bigr)\frac{d\psi}{dt}\vec X\times\vec e_{\psi}.
\end{equation}

Its total change as the particle moves from $\psi_1$ to
$\psi_2$ (it is supposed that this change is little and the vector $\vec X$ is originally oriented toward
 $\psi=0$ ) is:

\begin{equation} \label{ex7}
\Delta\alpha=\int\limits^{\psi_2}_{\psi_1}\omega dt=\frac{3\gamma MJ^2}{mc^2}
\int\limits^{\psi_2}_{\psi_1}\frac{\cos\psi d\psi}{Xr^2}.
\end{equation}

If $\vec X$ is constant and oriented toward $\psi=0$ we have

\begin{equation} \label{ex8}
\vec X\vec r=Xr\cos\psi=J^2-\gamma Mmr.
\end{equation}

From the unperturbed orbit (\ref{ex8}) we express $r$ and
substitute into (\ref{ex7}).
For a bound orbits ($m\neq 0$) with the eccentricity $e=A/M$, and the semi-major axis $a=J^2/\gamma Mm^2(1-e^2)$
we find the perihelion
precession:
\begin{eqnarray} \label{ex9}
\Delta\alpha=\frac{3\gamma Mm}{c^2J^2}
\int\limits^{2\pi}_{0}\frac{(X\cos\psi+\gamma Mm)^2}{X}\cos\psi d\psi= \nonumber \\
=\frac{6\pi\gamma^2m^2M^2}{c^2J^2}=\frac{6\pi\gamma M}{c^2a(1-e^2)}.  \  \  \  \  \
\end{eqnarray}

The perihelion shift of Mercury is equal to $\Delta\alpha=43^{\prime\prime}$ per century.

To calculate the deflection of light in the gravitational
field it is necessary to write the Lagrangian (\ref{ex2}) without an assumption of small velocities, so it can be
used in the ultra-relativistic case. Using the formalism \cite{5,6} this Lagrangian can be written in the form

\begin{eqnarray}
L=\frac{m_1 m_2}{r_{21}}[1+2f(\eta^2)\beta^2],
\end{eqnarray}

where $\eta^2=(r\times\beta)^2$, $\beta=v/c$.
The function $f$ is defined in \cite{5}:

\begin{equation}
f(x) = \frac{1}{1+\sqrt{1-x}}\simeq\frac{1}{2}+\frac{1}{8}x+...
\end{equation}

The expression for the energy of particlesmoving with
the speed of light (photons) has the form (\ref{ex3}) but
excluding the Newtonian interaction. The last term
in (\ref{ex3}) can be written as
 $\frac{\gamma MJ^2}{\varepsilon r^3}$,
where $\varepsilon$ is the photon frequency.
 Therefore for an unbound orbit, with the
photon mass $m=0$ we have:

\begin{equation} \label{ex13}
\Delta\alpha=\frac{3\gamma M\varepsilon}{c^4J^2}
\int\limits^{\pi/2}_{-\pi/2}X\cos^3\psi d\psi=\frac{4\gamma M}{c^2b},
\end{equation}

where $b=\frac{\varepsilon J^2}{Ac^2}$ is the parameter. Therefore for a ray
passing by the edge of the Sun, $\Delta\alpha=1,75^{\prime\prime}$.

Let us now calculate the radar echo delay. To do
that, let us integrate once more the obtained expression
in $dr$, taking into accou nt only
$\Delta t=\frac{2\gamma M\varepsilon X}{c^5J^2}
\int\frac{r_0}{r}dr$, where we have substituted $sin\psi=r_0/r$.
Here it is necessary to include the maximum delay time during
the signal motion there and back. As a result, we obtain

\begin{equation}
2\Delta t=\frac{4\gamma
M}{c^3}\ln(\frac{4r_Mr_Z}{r_0^2})\sim 240 mks,
\end{equation}

where $r_0=b$ is approximately equal to the Sun radius, $r_M$ and $r_Z$ are the distance from Mercury and
from Earth to the Sun, respectively. These results
for the anomalous perihelion procession and radar echo delay, obtained in the framework of the vector
theory of gravity, coincide with the analogous results
of general relativity
 \cite{3,7} and have been confirmed by
experiments \cite{1}.
The same results can also be obtained by means
of an effective geometrization of the Lagrangian (\ref{ex2}).
 Let us rewrite it in the following form:

\begin{equation} \label{ef1}
L=-mc^2(1-v^2/c^2)^{1/2}-m\varphi-m\varphi v^2/c^2
\end{equation}

In general relativity the Lagrangian leading to the
geodesic equation is written in the form \cite{1}:

\begin{equation}
L = -mc^2\Bigl(-g_{ik}\frac{dx^i}{dt}\frac{dx^k}{dt}\Bigr)^{1/2}.
\end{equation}

We write the metric tensor $g_{ik}$ in the form $g_{ik}=g^{0}_{ik}+h_{ik}$,
where $g^{0}_{ik}$ is the Minkowski metric and $h_{ik}$ are
corrections describing the gravitational field. Then
the Lagrangian takes the form

\begin{equation} \label{ef3}
L = -mc^2(1-v^2/c^2-h_{00}-2h_{0j}v^j-h_{jk}v^jv^k)^{1/2},
\end{equation}

where $j,k = 1,2,3$. Expanding the expression under
the square root and comparing (\ref{ef1}) and (\ref{ef3}) the metric $g_{ik}$ can be found up to second-order terms,
which corresponds to the post-Newtonian approximation:

\begin{eqnarray} \label{ef4}
g_{00}=-1-2\varphi  \nonumber,  \\
g_{\alpha\alpha}=1-2\varphi,  \nonumber  \\
g_{0\alpha}=0. \nonumber
\end{eqnarray}

Solving the Hamilton–Jacobi equation, we find, on
the basis of the derived metric tensor $g_{ik}$
 Mercury's perihelion shift, light deflection and radio echo delay.
As a result, these values coincide with the experimental
ones \cite{1}.

\section{Cosmology}
As is known, the cosmological Friedmann solution
can be derived in the framework of Newton's
theory \cite{8}.
We deduce the cosmological equation of
vector gravity theory neglecting the vector potential.

There is a theorem according to which the substance
surrounding a certain region with a spherically
symmetric layer does not affect the processes in this
region in any way. This statement holds true for a
region filled with a substance with constant density
in an infinite space. The theorem is true both for
Newton's theory and general relativity \cite{8},
 and as well for vector gravity theory.
Let us consider a spherical region of radius  $a$, inside which a substance with a density  $\rho$
has, at the time instant
$t=t_0$ a velocity distributed according to the law

\begin{equation}\label{400}
\vec{u}=H\vec{r}.
\end{equation}

The particle acceleration at a radius $a$ is

\begin{equation}\label{40}
\frac{du_a}{dt}=\frac{d^2a}{dt^2}=-\gamma\frac{M}{a^2},
\end{equation}

where $M=\frac{4\pi}{3}\rho a^3$ is the mass of matter inside
the relevant spherical region. To integrate the equation,
we multiply its both sides by
$u_a=da/dt$ and obtain

\begin{equation}\label{41}
\frac{1}{2}\Bigl(\frac{da}{dt}\Bigr)^2-\frac{4\pi\gamma}{3}\rho a^2=const.
\end{equation}

Eq. (\ref{40}) may be rewritten as

\begin{equation}\label{42}
\frac{du_a}{dt}=\frac{d^2a}{dt^2}=-\gamma\frac{4\pi}{3}\Bigl(\rho+\frac{3p}{c^2}\Bigr)a.
\end{equation}

taking into account the relation between changes in
the energy density $\varepsilon$ and the pressure $p$ \cite{8} and \cite{41}
can be integrated as follows:

\begin{equation}\label{43}
\frac{1}{2}H^2a^2-\frac{4\pi\gamma}{3}\rho a^2=const.
\end{equation}

From here we obtain the critical density value at given
H: $\rho_c=3H^2/8\pi\gamma$ where $H=\frac{1}{a}\frac{da}{dt}=
\frac{2}{3}\frac{1}{t}$ is the Hubble parameter. The constant may
be redenoted as $const=k*const^{'}$,
where $k=0,-1,+1$, and $const^{'}> 0$ which corresponds to flat,
closed and hyperbolic Universe models.

Eqs (\ref{40}) or (\ref{42}) and (\ref{41}) are equivalent to the
Friedmann equation, which, for the case of zero pressure
($p = 0$) and ultrarelativistic matter ($p=\varepsilon/3$),
their solutions are given, for example, in \cite{7,8}.

At model construction, a region with a certain
amount ofmatter $M$ and a certain $a(t)$ we considered.
However, the results for the quantities $\rho(t)$ and $H(t)$ turned out to be independent of the choices of $M$ and $a$
 that confirms the possibility of extending the calculation
to infinite space. Sometimes, a gravitational
paradox is discussed for Newton's theory. However,
there is a consequent method of reasoning that does
not lead to a paradox. Consider a sphere of final size $a$ with a certain density $\rho$
 and velocity profile $\vec{u}=H\vec{r}$.
A solution of the mechanical problem for it leads to
a certain relation for $H(t)$ and $\rho(t)$, not including $a$.
Therefore, if $a\to\infty$ at some $t_0$ with fixed $H(t_0)$ and $\rho(t_0)$, then a correct solution is found for an
infinite homogeneous Universe.

So far it was assumed that there is a selected fixed
point at the center of a spherical region. At any other
point, matter is moving with a certain velocity, and
there is a preferred direction specified by the velocity
vector $\vec{u}$.
 However, it is easy to reveal that this
selection of the center and direction is only apparent.
Now take a random point $X$ inside the sphere and
pass on to the coordinate system (reference frame)
where this point is at rest. On the basis of the classical
transformation laws we obtain that $\vec{r}^{'}_{c}=\vec{r}_c-\vec{r}_x$,
 $\vec{u}^{'}_{c}=\vec{u}_c-\vec{u}_x$,
where the prime denotes variables of
in the new coordinates system. Substituting to the
Hubble law, we get $\vec{u}^{'}=H\vec{r}^{'}$.
The law of motion from the point of view of an observer at the point $X$ has
 no difference from that for an observer at the
assumed center of the sphere. Since the above region
is selected only mentally in the infinite homogeneous
matter distribution, the point $X$ is entirely equivalent
to the assumed center or any randomly chosen
point. Hence the solution constructed complies with
the homogeneity principle and is necessarily nonstationary.

Following  \cite{8} let us consider the cosmological redshift.
In our case it is determined by the Doppler
effect and motion in a gravity force field.
Up to the second order, the redshift expression
related to the Doppler effect, which manifests itself in
a wavelength increase caused by the expansion of the
universe, is as follows:

\begin{equation}
 1-z=\frac{\omega^{*}}{\omega}=1-\frac{v}{c}+\frac{1}{2}\Bigl(\frac{v}{c}\Bigr)^2,
\end{equation}

 where $\omega$ - is the source frequency, $\omega^{*}$ - is the light
frequency in the laboratory system moving with a
velocity $v$. To acquire a complete redshift, one should
also consider the motion in the gravitational field:

\begin{equation}
 \omega-\omega^{*}=\frac{\varphi(a)}{\varphi(0)}{c^2}\omega^{*}
\end{equation}

  where $\varphi$ is the Newtonian gravity potential and $a$ is
the distance from the emission point to the origin.
For a universe with zero pressure
    $p=0$, the matter density  $\rho_0$ and the Hubble velocity field
 $v=H_0a$ at
a given moment $t_0=0$, expanding the hydrodynamic
equations of continuity and motion in a series according
with respect to the small parameter $t$, we obtain:

 \begin{eqnarray}\label{44}
\rho=\rho_0-3H_0t\rho_0; \nonumber \\
v=H_0a-H^2_0\Bigl(1+\frac{\Omega}{2}\Bigr)at,
 \end{eqnarray}

where $\Omega=\rho_0/\rho_c$, $\rho_c=\frac{3H_0^2}{8\pi\gamma}$ is the critical
density. Thus, in the present approximation, using
the expression for the velocity (\ref{44}) we obtain an
expression for the cosmological redshift:

\begin{equation}
z=\frac{H_0}{c}a+\frac{H_0^2}{c^2}a^2\Bigl(\frac{1}{2}+\frac{\Omega}{4}\Bigr).
\end{equation}

Solving this equation with the required accuracy, we
obtain:

\begin{eqnarray}
a=\frac{c}{H_0}\Bigl[z-\Bigl(\frac{1}{2}+\frac{\Omega}{4}\Bigr)z^2\Bigr], \\
t=-\frac{1}{H_0}\Bigl[z-\Bigl(\frac{1}{2}+\frac{\Omega}{4}\Bigr)z^2\Bigr].
\end{eqnarray}

The redshift value is a function of the parameters $a$ and $\Omega$, where $a$ is the distance from the source at time
t of light emission, which is received at time  $t=0$
with the redshift $z$ by an observer located at the origin.
This expression coincides with the corresponding relation
of GR in the second approximation \cite{7,8}.

\section{Conclusion}

We have studied amodel in which the gravitational
interaction is described by a 4-component vector potential.

We have calculated the values of theMercury orbit
perihelion shift, the light deflection angle in the gravitational
field of the Sun and the radar echo delay in
a post-Newtonian approximation. The values found
coincide with the experimental ones.

It has been shown that, in the framework of this
theory, there exist models of an expanding Universe.


\end{document}